\documentclass[journal, letterpaper]{IEEEtran}
\IEEEoverridecommandlockouts

\usepackage{algorithm}
\usepackage{comment}
\usepackage{hyperref}
\usepackage{subcaption}

\usepackage{cite}
\usepackage{amsmath,amssymb,amsfonts}
\usepackage{algorithmic}
\usepackage{graphicx}
\usepackage{textcomp}
\usepackage{xcolor}
\def\BibTeX{{\rm B\kern-.05em{\sc i\kern-.025em b}\kern-.08em
    T\kern-.1667em\lower.7ex\hbox{E}\kern-.125emX}}

\begin{document}
\title{Fairness-Aware Secure Communication in ISAC Systems with STAR-RIS and RSMA
\thanks{This work was accepted and presented at the 21st International Conference on Wireless and Mobile Computing, Networking and Communications (WiMob 2025) in Marrakech, Morocco, on 22 October 2025.

Thanh Nha To is with the Department of Telecommunications Engineering, Ho Chi Minh City University of Technology (HCMUT), Vietnam National University Ho Chi Minh City (VNU-HCM), and also with the Wireless Broadband R\&D Center, Viettel High Technology Industries Corporation (VHT), Viettel Group, Vietnam (emails: tothanhnha1401@gmail.com, nhatt30@viettel.com.vn).

Hoang Lai Pham, Quynh Nguyen Thi, Tuan Anh Pham, and Le Thanh Bang are with the Wireless Broadband R\&D Center, Viettel High Technology Industries Corporation (VHT), Viettel Group, Vietnam (email: \{laiph3, quynhnt57, tuanpa44,
banglt2\}@viettel.com.vn).
}
}

\author{
\IEEEauthorblockN{
Thanh Nha To,
Hoang Lai Pham,
Quynh Nguyen Thi, 
Tuan Anh Pham, and
Le Thanh Bang} 
}

\maketitle

\begin{abstract}
In this paper, we investigate the integration of simultaneously transmitting and reflecting reconfigurable intelligent surfaces (STAR-RIS) with rate-splitting multiple access (RSMA) for improving physical layer security (PLS) in integrated sensing and communication (ISAC) systems. Specifically, we consider a multi-user, multi-sensing-target scenario, where each sensing target is treated as a potential eavesdropper, reflecting realistic deployment conditions. To enable fairness-aware secure communication among users while maintaining sensing performance, we formulate a joint optimization problem that designs the base station beamforming vectors and STAR-RIS coefficients, aiming to maximize the minimum secrecy rate under a minimum beampattern gain constraint. To solve the resulting non-convex problem, we propose an efficient algorithm based on alternating optimization (AO) and the majorization–minimization (MM) method. Simulation results verify the fast convergence of the proposed algorithm and demonstrate significant improvements in secure communication performance.
\end{abstract}


\section{Introduction}
Integrated sensing and communication (ISAC) has gained traction as a key enabler of sixth-generation (6G) wireless networks, aiming to extend the functionality of wireless systems beyond data transmission to support diverse applications such as unmanned aerial vehicles, intelligent transportation systems, and the internet of things \cite{gonzalez2024integrated}. By integrating sensing and communication functionalities into a unified framework, ISAC can significantly improve spectral and energy efficiency while enabling new capabilities in environmental awareness and system intelligence. However, ISAC also poses noticeable challenges in ensuring physical layer security (PLS), particularly against eavesdropping attacks \cite{su2020secure} as adversaries may utilize sensing signals as a means to obtain private information. This critical security threats have drawn significant attention for seeking complementary solutions for enhancing both sensing and communication capacity of ISAC, in which simultaneously transmitting and reflecting reconfigurable intelligent surfaces (STAR-RIS) \cite{xu2021star} and artificial noise (AN) \cite{su2020secure} are widely adopted, often denoted as STAR-RIS-assisted secure ISAC system. STAR-RIS have emerged as a promising solution to further improve ISAC secrecy performance by enabling flexible electromagnetic wave manipulation in both transmission and reflection modes \cite{wei2024star}. The authors in \cite{liu2023exploiting} were among the first to explore the use of STAR-RIS for enhancing the security performance of ISAC systems with the presence of an eavesdropper, formulating an optimization problem that maximizes the secrecy rate while guaranteeing a minimum sensing signal to interference plus noise ratio (SINR). To address channel variations, \cite{zhu2024ai} proposed a coupled phase-shift STAR-RIS model employing both time switching and energy splitting operating protocols to ensure secure transmission in ISAC systems. In addition, \cite{kamal2025optimizing} extended the study to more practical multi-user, multi-eavesdropper scenarios,

Beyond the use of STAR-RIS, integrating Rate-splitting multiple access (RSMA) into ISAC systems has also emerged as a potential strategy to enhance PLS, owing to its superior interference management and flexible message splitting capabilities. As a recently proposed multiple access scheme, RSMA has shown great promise in balancing the communication and sensing trade-offs  in secure ISAC systems. Compared to traditional techniques such as Non-orthogonal multiple access (NOMA) \cite{wei2024star} and Space division multiple access (SDMA), RSMA offers enhanced spectral efficiency, user fairness, and robustness to imperfect channel state information \cite{zhang2024robust}. The paper \cite{zhang2024robust, salem2024robust} investigated a secure downlink RIS-aided RSMA system with multiple legitimate users, and a potential eavesdropper to improve ISAC system security. 


While both RSMA and STAR-RIS have individually demonstrated their benefits in ISAC systems, their combined potential for enhancing PLS has not been fully explored. Most existing works either disregard security considerations \cite{liu2025star,chen2025joint} or simplify the system to single-user or single-target scenarios with conventional RIS rather than STAR-RIS \cite{zhang2024robust, salem2024robust}.  In \cite{liu2025star,chen2025joint}, a STAR-RIS-enabled downlink ISAC system with RSMA transmission was taken into account, where the base station (BS) jointly served multiple users and a sensing target. While \cite{liu2025star} focused on maximizing the sensing SINR, \cite{chen2025joint} aimed at maximizing the achievable sum rate. However, neither work considered the security aspect, leaving the PLS of STAR-RIS-enabled RSMA-ISAC systems an open research problem. The joint optimization of RSMA-assisted STAR-RIS ISAC for secure communication and accurate sensing in realistic multi-user, multi-target settings remains largely unaddressed.

In this paper, we investigate the integration of STAR-RIS into RSMA-assisted ISAC systems for optimizing secure communication in the presence of multiple sensing targets. The key contributions of this paper are as follows:

\begin{itemize}
    \item A STAR-RIS-assisted RSMA-ISAC framework is introduced to enhance PLS while satisfying sensing requirements in multi-user, multi-target scenarios, where each sensing target is considered a potential eavesdropper.
    \item A joint optimization problem is formulated to maximize the minimum secrecy rate across all communication users, subject to a minimum beampattern gain constraint and BS power limitations.
    \item To solve the resulting non-convex problem, an efficient algorithm based on AO and MM approaches is proposed.
    \item Simulation results verify that the proposed scheme effectively achieves a favorable trade-off between secure communication and sensing performance, outperforming existing baseline methods. 
\end{itemize}

\section{System Model and Problem Formulation}
\subsection{System Model}

\begin{figure}[htbp]
\centering
\includegraphics[width=1\linewidth]{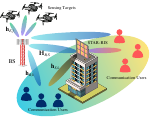}
\caption{Secure ISAC system model supported by STAR-RIS.}
\label{fig:sys-model}
\end{figure}

As shown in Fig. \ref{fig:sys-model}, we consider a secure RSMA-ISAC system assisted by a STAR-RIS. The system consists of a BS, denoted by $B$, equipped with $N_B$ antennas arranged in a uniform linear array (ULA), and a STAR-RIS, denoted by $S$, comprising $N_S$ elements configured as a uniform planar array (UPA). The transmission and reflection coefficients of the STAR-RIS are denoted by $\mathbf{c}_t$ and $\mathbf{c}_r$, respectively, and the set of STAR-RIS elements is represented by $\mathcal{N}=\{1, 2, \dots, N_S \}$. The system serves $K_c$ communication users (CUs) located on both sides of the STAR-RIS, where $\mathcal{C} = \{1, 2, \dots, K_c\}$ denotes the set of CUs. Additionally, the BS serves $K_s$ sensing targets (SUs) located in its coverage region, represented by the set $\mathcal{S} = \{1, 2, ..., K_s \}$. Each CU and SU is assumed to be equipped with a single receive antenna. Let $\mathcal{T}$ and $\mathcal{R}$ denote the sets of CUs located in the transmission and reflection regions of the STAR-RIS, respectively.

The BS transmits a common stream $x_{c} \in \mathbb{C}$ and a private stream $x_{p, k} \in \mathbb{C}$ to the $k$-th CU $(k \in \mathcal{C})$ via the beamforming vectors $\mathbf{w}_c$ and $\mathbf{w}_{p, k} \in \mathbb{C}^{N_B}$, respectively. Let $\mathbf{W} = [\mathbf{w}_c, \mathbf{w}_{p, 1}, \mathbf{w}_{p, 2} , \dots, \mathbf{w}_{p, K_{c}}]$ represent the set of beamforming vectors for the common and private streams. In addition, to simultaneously reinforce sensing performance and improve the secrecy rate by interfering potential eavesdroppers, the BS also transmits AN $\mathbf{x}_s \in \mathbb{C}^{N_B}$ via a beamforming matrix $\mathbf{W}_s \in \mathbb{C}^{N_B \times N_B}$. The transmitted signal from the BS is given by $\mathbf{s} =  \mathbf{W}_s \mathbf{x}_s + \mathbf{w}_c x_c + \sum_{k \in \mathcal{C}} \mathbf{w}_{p, k} x_{p, k}$, where the transmitted signals are assumed to be mutually independent and normalized to unit power. Letting $\mathbf{R}_s = \mathbf{W}_s \mathbf{W}_s^H$, the total transmit power at the BS is expressed as
\begin{align}
P_B  &= \mathrm{tr}( \mathbf{W}_s \mathbf{W}_s^H) +  \mathrm{tr}( \mathbf{w}_c \mathbf{w}_c^H ) + \sum_{k \in \mathcal{C}}  \mathrm{tr}( \mathbf{w}_{p, k} \mathbf{w}_{p, k} ^H ) \nonumber \\
 &= \mathrm{tr}( \mathbf{R}_s) +  \|\mathbf{w}_c \|^2 + \sum_{k \in \mathcal{C}} \| \mathbf{w}_{p, k} \|^2. 
\end{align}

The channel from the BS to the STAR-RIS is modeled by the MIMO matrix \( \mathbf{H}_{B,S} \in \mathbb{C}^{N_S \times N_B} \). The direct channel between the BS and the \( k \)-th CU is denoted by \( \mathbf{h}_{B,k} \in \mathbb{C}^{N_B} \), while the channel from the STAR-RIS to user is denoted by \( \mathbf{h}_{S,k} \in \mathbb{C}^{N_S} \). For the SUs, the channel between the BS and the \( j \)-th SU \( (j \in \mathcal{S}) \), is denoted by \( \mathbf{h}_{E,j} \in \mathbb{C}^{N_B} \), which is assumed to follow a line-of-sight model. 
\subsection{Communication Model}
The received signal at CU $k \in \mathcal{C}$ is given as follows:
\begin{align}
y_k =
\begin{cases}
\left(\mathbf{h}_{B, k}^H + \mathbf{h}_{S, k}^H \mathrm{diag}(\mathbf{c}_t) \mathbf{H}_{B, S}\right) \mathbf{s} + n_{k} , & k \in \mathcal{T}, \\
\left(\mathbf{h}_{B, k}^H + \mathbf{h}_{S, k}^H \mathrm{diag}(\mathbf{c}_r) \mathbf{H}_{B, S}\right) \mathbf{s} + n_{k} , & k \in \mathcal{R},
\end{cases}
\end{align}
where $n_{k} $ is additive white Gaussian noise (AWGN) with zero mean and variance $\sigma_{c, k}^2$. 
For notational simplification, the received signal for a reflection-side user \( k \in \mathcal{R} \) can be equivalently expressed as:
\begin{align}
y_k &= \left(\mathbf{h}_{B, k}^H + \mathbf{c}_r^T \mathrm{diag}(\mathbf{h}_{S, k})^H \mathbf{H}_{B, S}\right) \mathbf{s} + n_{k} \nonumber \\
&=\begin{bmatrix}
\mathbf{c}_r \\
1
\end{bmatrix}^T
\begin{bmatrix}
\mathrm{diag}(\mathbf{h}_{S, k})^H \mathbf{H}_{B, S} \\
\mathbf{h}_{B, k}^H 
\end{bmatrix}
\mathbf{s} + n_k \nonumber \\
&= \sigma_{c, k} \mathbf{v}_r^H \mathbf{G}_k \mathbf{s} + n_k, \quad k \in \mathcal{R},
\end{align}
where $\mathbf{G}_k =\sigma_{c, k} ^{-1} [\mathbf{H}_{B, S}^H\mathrm{diag}(\mathbf{h}_{S, k}), \,  \mathbf{h}_{B, k}]^H, \, \forall k \in \mathcal{C} $. Let $\mathbf{v}_k = \mathbf{v}_r = [\mathbf{c}_r^T, \, 1]^H$, if $k \in \mathcal{R}$, and $\mathbf{v}_k = \mathbf{v}_t = [\mathbf{c}_t^T, \, 1]^H$ otherwise. Set $\mathbf{V} = [\mathbf{v}_t, \mathbf{v}_r]$.

The achievable rate of the common stream at CU  $k \in \mathcal{C}$ is given by the Shannon capacity formula as 
\begin{align}
& \quad R_{c, k} \nonumber \\
&= \log_2 \left(1 + \frac{\left| \mathbf{v}_k^H \mathbf{G}_k \mathbf{w}_c \right|^2}{\|\mathbf{v}_k^H \mathbf{G}_k  \mathbf{W}_s\|^2 + \sum_{i \in \mathcal{C}} |\mathbf{v}_k^H \mathbf{G}_k \mathbf{w}_{p, i} |^2 + 1 } \right) \nonumber \\
&= \log_2 \left(1 + \frac{\left| \mathbf{v}_k^H \mathbf{G}_k \mathbf{w}_c \right|^2}{E_{c, k} (\mathbf{R}_s, \mathbf{W}, \mathbf{V}) - \left| \mathbf{v}_k^H \mathbf{G}_k \mathbf{w}_c \right|^2 } \right),
\end{align}
where $E_{c, k} (\mathbf{R}_s, \mathbf{W}, \mathbf{V}) = \mathbf{v}_k^H \mathbf{G}_k  \mathbf{R}_s \mathbf{G}_k^H \mathbf{v}_k + |\mathbf{v}_k^H \mathbf{G}_k \mathbf{w}_{c} |^2 + \sum_{i \in \mathcal{C}} |\mathbf{v}_k^H \mathbf{G}_k \mathbf{w}_{p, i} |^2 + 1, k \in \mathcal{C} $.
Under the RSMA framework, each CU decodes the private stream after applying successive interference cancellation (SIC) to remove the common stream. The achievable rate of private stream for CU $k \in \mathcal{C}$ is 
\begin{align}
& \quad R_{p, k} \nonumber \\
&= \log_2 \left(1 + \frac{\left| \mathbf{v}_k^H \mathbf{G}_k \mathbf{w}_{p, k} \right|^2}{\|\mathbf{v}_k^H \mathbf{G}_k  \mathbf{W}_s\|^2 + \sum_{i \in \mathcal{C} \setminus \{k\}} |\mathbf{v}_k^H \mathbf{G}_k \mathbf{w}_{p, i} |^2 + 1 } \right) \nonumber \\
& =\log_2 \left(1 + \frac{\left| \mathbf{v}_k^H \mathbf{G}_k \mathbf{w}_{p, k} \right|^2}{E_{p, k} (\mathbf{R}_s, \mathbf{W}, \mathbf{V}) - \left| \mathbf{v}_k^H \mathbf{G}_k \mathbf{w}_{p, k} \right|^2 } \right)
\end{align} 
where the term $E_{p, k} (\mathbf{R}_s, \mathbf{W}, \mathbf{V}) = \mathbf{v}_k^H \mathbf{G}_k  \mathbf{R}_s \mathbf{G}_k^H \mathbf{v}_k  + \sum_{i \in \mathcal{C}} |\mathbf{v}_k^H \mathbf{G}_k \mathbf{w}_{p, i} |^2 + 1,  k \in \mathcal{C}$.

Next, we consider the security aspect of the system. Each SU $j \in \mathcal{S}$, potentially acting as an eavesdropper, may overhear the common stream $x_c$ with an eavesdropping rate given by 
\begin{align}
R_{c}^{e, j} = \log_2 \left(1 + \frac{\left| \mathbf{g}_j^H \mathbf{w}_c \right|^2}{D_{j} (\mathbf{R}_s, \mathbf{W}) - \left| \mathbf{g}_j^H \mathbf{w}_c \right|^2 } \right), \label{eq: Rce}
\end{align}
where the term $D_{j} (\mathbf{R}_s, \mathbf{W}) = \mathbf{g}_j^H \mathbf{R}_s \mathbf{g}_j + |\mathbf{g}_j^H \mathbf{w}_{c} |^2 + \sum_{i \in \mathcal{C}} |\mathbf{g}_j^H \mathbf{w}_{p, i} |^2 + 1, j \in \mathcal{S} $, $\mathbf{g}_j = \sigma_{s, j} ^{-1} \mathbf{h}_{E, j} $, and $\sigma_{s, j}^2$ denotes the noise power at sensing target $j$. As SUs are unable to decode or cancel the common stream $x_c$, the eavesdropping rate for the private stream $x_{p, k} $ at target $j \in \mathcal{S}$ is 
\begin{align}
R_{p, k}^{e, j} = \log_2 \left(1 + \frac{\left| \mathbf{g}_j^H \mathbf{w}_{p, k} \right|^2}{D_{j} (\mathbf{R}_s, \mathbf{W}) - \left| \mathbf{g}_j^H \mathbf{w}_{p, k} \right|^2 } \right), \label{eq: Rpe}
\end{align}
According to \cite{wyner1975wire}, the secrecy rates of the common and private streams are defined as $R_{c, k}^{sec} = [R_{c, k} - \max_{j \in \mathcal{S}} R_{c}^{e, j}]^+ $, $R_{p, k}^{sec} = [R_{p, k} - \max_{j \in \mathcal{S}} R_{p, k}^{e, j}]^+ $, respectively, where $[z]^+ = \max(0, z)$. To ensure secure transmission to CU $k$, the actual private rate must not exceed $R_{p, k}^{sec}$, and the total allocated common rate $\sum_{k \in \mathcal{C}} r_k$ is limited by $\min_{k \in \mathcal{C}} R_{c, k}^{sec}$, where $r_k$ is the portion of the common rate allocated to CU $k$. Therefore, the total achievable secrecy rate for CU $k$ is 

\begin{align}
R_{k} &= r_{k} + R_{p, k}^{sec}.
\end{align}

\subsection{Sensing Model}
To evaluate the sensing performance, we consider the beampattern gain received at the sensing target $j \in \mathcal{S}$, which is defined as
\begin{align}
G_j & = \mathbb{E} \left\{ \left| \mathbf{a}_j^H \mathbf{s} \right|^2 \right\}   \nonumber \\
& = \mathbf{a}_j^H \left( \mathbf{R}_s + \mathbf{w}_c \mathbf{w}_c^H + \sum_{i \in \mathcal{C}} \mathbf{w}_{p, i} \mathbf{w}_{p, i} ^H \right) \mathbf{a}_j.
\end{align}
where, $\mathbf{a}_j = [1, e^{j2\pi\sin \theta_j\Delta d/\lambda}, \dots, e^{j\pi2\sin \theta_j\Delta d/\lambda (N_B-1)} ]^T$ denotes the steering vector at angle $\theta_j$ of SU $j$, with $\lambda$ and $\Delta d$ denoting the carrier wavelength and the spacing between two adjacent antennas, respectively. In the sensing-only case, where no communication constraints are considered, we address the following utility maximization problem
\begin{align}
\mathbf{R}^{{opt}} 
= \arg \max_{\mathbf{R} \succeq 0}  
\min_{j \in \mathcal{S}} 
\mathbf{a}_j^H \mathbf{R} \mathbf{a}_j 
\quad 
\text{s.t.} \quad 
\mathrm{tr} ( \mathbf{R} ) \leq P_B^{max} .
\end{align}
Here, \(P_B^{{max}}\) denotes the maximum transmit power budget at the BS. The optimal beampattern gain at  SU $j$ is defined as $G_j^{opt} = \mathbf{a}_j^H \mathbf{R}^{opt} \mathbf{a}_j$.

\subsection{Problem Formulation}
The objective of this study is to jointly optimize the beamforming vectors at the BS and the coefficients of the STAR-RIS in order to maximize the minimum secrecy rate of all CUs. The optimization problem is subject to the following constraints: secrecy rate requirements for each CU, minimum beampattern gain for each SU, a total transmit power budget at the BS, and the energy conservation law inherent to STAR-RIS operation. The problem can be formulated as follows:
\begin{subequations}
\label{eq:P1}
\begin{align}
\max_{\mathbf{R}_s, \mathbf{W}, \mathbf{V}, \mathbf{r}} \quad & \min_{k \in \mathcal{C}} R_k \label{eq:P1-obj} \\
\text{s.t.} \qquad
& R_{p, k}^{sec} \geq 0, \quad \forall k \in \mathcal{C}, \label{eq:P1-sec-p} \\
& \sum_{i \in \mathcal{C}} r_{i} \leq  R_{c, k}^{sec}, \quad \forall k \in \mathcal{C},  \label{eq:P1-sec-c} \\
& {r}_k \geq {0}, \quad \forall k \in \mathcal{C},  \label{eq:P1-r-c} \\
& G_j \geq \eta_j G_j^{opt}, \quad \forall j \in \mathcal{S}, \label{eq:P1-g} \\
& P_B \leq P_B^{{max}}, \label{eq:P1-p}\\
& |v_{t, n} |^2 + |v_{r, n} |^2 \leq 1, \quad \forall n \in \mathcal{N} , \label{eq:P1-c}  \\
& v_{t, N_S+1} = v_{r, N_S + 1} = 1. \label{eq:P1-c-1}
\end{align}
\end{subequations}
where $\eta_j \in (0, 1)$ denotes the ratio of the beampattern gain relative to $G_j^{opt} $. 

It is straightforward to observe that the constraints \eqref{eq:P1-r-c}, \eqref{eq:P1-p}, \eqref{eq:P1-c} and \eqref{eq:P1-c-1} are convex. Nevertheless, the objective function in \eqref{eq:P1-obj} and the others are non-convex. As a result, the formulated problem is a highly non-convex optimization problem, which poses significant challenges for obtaining a globally optimal solution.

\section{Proposed Iterative Algorithm Based on AO}
\subsection{Problem Reformulation} \label{sec: problem2-refo}
To address the problem, we introduce auxiliary variable sets $\alpha = \{\alpha_c, \alpha_{p, 1}, \dots, \alpha_{p, K_c} \}$, $\beta = \{\beta_c, \beta_{p, 1}, \dots, \beta_{p, K_c} \}$, and $\omega$. Hence, the original problem \eqref{eq:P1} is equivalently transformed as follows 
\begin{subequations}
\label{eq:P2}
\begin{align}
\max_{\mathbf{R}_s, \mathbf{W}, \mathbf{V}, \mathbf{r}, \alpha, \beta, \omega} \quad & \omega \\
\text{s.t.} \quad \quad \quad
& R_{c, k} \geq \alpha_c, \quad \forall k \in \mathcal{C}, \label{eq:P2-a-c} \\
& R_{p, k} \geq \alpha_{p, k} , \quad \forall k \in \mathcal{C}, \label{eq:P1-a-p} \\
& R_{c}^{e, j} \leq \beta_{c}, \quad \forall j \in \mathcal{S}, \label{eq:P2-b-c} \\
& R_{p, k}^{e, j} \leq \beta_{p, k} , \quad \forall k \in \mathcal{C}, \forall j \in \mathcal{S}, \label{eq:P2-b-p} \\
& r_k + \alpha_{p, k} - \beta_{p, k} \geq \omega, \quad \forall k \in \mathcal{C}, \label{eq:P2-r} \\
& \alpha_{p, k} - \beta_{p, k} \geq 0 , \quad k \in \mathcal{C}, \label{eq:P2-r-p} \\
& \sum_{k \in \mathcal{C}} r_k \leq \alpha_c - \beta_c, \label{eq:P2-r-c} \\
& \eqref{eq:P1-r-c}, \eqref{eq:P1-g}, \eqref{eq:P1-p}, \eqref{eq:P1-c}, \eqref{eq:P1-c-1}.
\end{align}
\end{subequations}

Based on the proportional relationship between $\mathbf{a}_j$ and $\mathbf{g}_j$, we let $D_j^{opt} = \mathbf{g}_j^H \mathbf{R}^{opt} \mathbf{g}_j $ and introduce the auxiliary variable set $\delta = \{\delta_1, \delta_2, \dots, \delta_{K_s} \}$, which is used to represent the beampattern gain constraints given in \eqref{eq:P1-g} as follows:
\begin{align}
& D_j (\mathbf{R}_s, \mathbf{W}) \geq \delta_j, \quad \forall j \in \mathcal{S}, \label{eq:P3-delta} \\
&\delta_j  \geq 1 +  \eta_j D_j^{opt}, \quad \forall j \in \mathcal{S}. \label{eq:P3-g}
\end{align}
With the variable $\delta_j$ satisfying the constraint \eqref{eq:P3-delta}, the eavesdropping rate corresponding to the common stream \eqref{eq: Rce} admits the following upper bound 
\begin{align}
R_{c}^{e, j} & \leq \log_2 \left(1 + \frac{|\mathbf{g}_{j}^H \mathbf{w}_{c}|^2}{ \delta_j - |\mathbf{g}_{j}^H \mathbf{w}_{c}|^2 } \right) \nonumber \\
& = \log_2 \delta_j  -  \log_2 \left|{\delta_j - |\mathbf{g}_{j}^H \mathbf{w}_{c}|^2} \right|, \quad \forall j \in \mathcal{S}.
\end{align}   
Likewise, the eavesdropping rate corresponding to the private stream \eqref{eq: Rpe} can be equivalently transformed 
\begin{align}
& R_{p, k}^{e, j} \leq \log_2 \left(1 + \frac{|\mathbf{g}_{j}^H \mathbf{w}_{p, k}|^2}{ \delta_j - |\mathbf{g}_{j}^H \mathbf{w}_{p, k}|^2 } \right) = \log_2 \delta_j \nonumber \\
& \qquad -  \log_2 \left|{\delta_j - |\mathbf{g}_{j}^H \mathbf{w}_{p, k}|^2} \right|, \quad \forall j \in \mathcal{S}, \forall k \in \mathcal{C}.
\end{align}  
Consequently, by introducing auxiliary variable set $\mu = \{\mu_1, \mu_2, \dots, \mu_{K_s} \}$, the eavesdropping rate constraint \eqref{eq:P2-b-c} and \eqref{eq:P2-b-p} can be equivalently rewritten as follows:
\begin{align}
& \log_2 \delta_j \leq \mu_j, \quad \forall j \in \mathcal{S}, \label{eq:P3-mu} \\
& \mu_j - \log_2 ( \delta_j - |\mathbf{g}_{j}^H \mathbf{w}_{c}|^2) \leq \beta_{c}, \quad \forall j \in \mathcal{S}, \label{eq:P3-b-c} \\
& \mu_j - \log_2 (\delta_j - |\mathbf{g}_{j}^H \mathbf{w}_{p, k}|^2) \leq \beta_{p, k}, \quad \forall j \in \mathcal{S}, \forall k \in \mathcal{C}. \label{eq:P3-b-p}
\end{align}    
 
By substituting these transformations, the problem \eqref{eq:P2} can thus be equivalently rewritten as the following optimization:
\begin{subequations}
\label{eq:P3}
\begin{align}
\max_{\mathbf{R}_s, \mathbf{W}, \mathbf{V}, \mathbf{r}, \alpha, \beta, \delta, \mu, \omega } \quad & \omega \\
\text{s.t.} \qquad \qquad 
& \eqref{eq:P2-a-c}, \eqref{eq:P1-a-p}, \eqref{eq:P3-mu}, \eqref{eq:P3-b-c}, \eqref{eq:P3-b-p},  \eqref{eq:P3-delta}, \\
& \eqref{eq:P3-g}, \eqref{eq:P2-r}, \eqref{eq:P2-r-p}, \eqref{eq:P2-r-c},  \\
& \eqref{eq:P1-r-c}, \eqref{eq:P1-p}, \eqref{eq:P1-c}, \eqref{eq:P1-c-1}.
\end{align}
\end{subequations}
\subsection{Optimization of BS Weights}
In this subsection, we adopt an AO-based approach to solve the reformulated problem \eqref{eq:P3}. At each iteration $\iota$, the variable $\overline{\mathbf{V}}$ is fixed, and the remaining variables are optimized accordingly. Next, the information rate constraint is lower bounded by a concave function using inequality of log function~\cite{niu2022joint}. The achievable rate of user \( k \) is lower bounded as follows:
\begin{align}
R_{m, k} & \geq f_{m, k} + 2 \Re \left (b_{m, k}^* {\mathbf{v}_k}^H \mathbf{G}_{k} \mathbf{w}_{c} \right) -  q_{c, k} {E}_{c, k} (\mathbf{R}_s, \mathbf{W}, {\mathbf{V}}) \nonumber \\
& \triangleq  \tilde{R}_{m, k} (\mathbf{R}_s, \mathbf{W}, {\mathbf{V}}) , \quad \forall k \in \mathcal{C}, m \in \{c, p \}, 
\end{align}  
where 
\begin{align}
    f_{m, k} &= \log_2 \left (\frac{ \overline{E}_{m, k} }{\overline{E}_{m, k} - |\overline{u}_{m, k}|^2 } \right)  - \frac{ |\overline{u}_{m, k}|^2 \log_2 e }{\overline{E}_{m, k} - |\overline{u}_{m, k}|^2 }, \\
    q_{m, k} &= \frac{\log_2 e}{\overline{E}_{m, k} - |\overline{u}_{m, k}|^2} - \frac{\log_2 e}{\overline{E}_{m, k}}, \\
    b_{m, k} &= \frac{\overline{u}_{m, k} \log_2 e}{\overline{E}_{m, k} - |\overline{u}_{m, k}|^2},
\end{align}
$\overline{E}_{m, k} = {E}_{m, k} (\overline{\mathbf{R}}_s, \overline{\mathbf{W}}, \overline{\mathbf{V}}), m \in \{c, p\}$, $\overline{u}_{c, k} = \overline{\mathbf{v}}_k^H \mathbf{G}_k \overline{\mathbf{w}}_c$, and  $\overline{u}_{p, k} = \overline{\mathbf{v}}_k^H \mathbf{G}_k \overline{\mathbf{w}}_{p, k} $.
During the optimization of BS weights, the rate constraints \eqref{eq:P2-a-c} and \eqref{eq:P1-a-p}  can be equivalently expressed in a convex form as follows
\begin{align}
\tilde{R}_{c, k}(\mathbf{R}_s, \mathbf{W}, \overline{\mathbf{V}}) & \geq \alpha_{c}, & \forall k \in \mathcal{C}, \label{eq:P3-w-a-c} \\
\tilde{R}_{p, k}(\mathbf{R}_s, \mathbf{W}, \overline{\mathbf{V}}) & \geq \alpha_{p, k}, & \forall k \in \mathcal{C}. \label{eq:P3-w-a-p}
\end{align}  

The beampattern gain constraint \eqref{eq:P3-delta} is transformed as
\begin{align}
D_j (\mathbf{R}_s, \mathbf{W}) & \geq \mathbf{g}_{j}^H \mathbf{R}_{s} \mathbf{g}_{j} + 2 \Re \left \{\overline{\mathbf{w}}_c^H 
 \mathbf{g}_j \mathbf{g}_j^H \mathbf{w}_c \right \} - | \mathbf{g}_j^H \overline{\mathbf{w}}_c |^2  \nonumber \\
 & + 2 \sum_{i \in \mathcal{C}}  \Re \left \{\overline{\mathbf{w}}_{p, i}^H \mathbf{g}_{j} \mathbf{g}_{j}^H \mathbf{w}_{p, i} \right \} - \sum_{i \in \mathcal{C}} | \mathbf{g}_{j}^H \overline{\mathbf{w}}_{p, i} |^2 + 1 \nonumber \\
 & \geq \delta_j , \quad \forall j \in \mathcal{S}. \label{eq:P3-w-delta}
\end{align}
Using inequality of log function, the eavesdropping rate constraint can be represented as
\begin{align}
\log_2 \left(\overline{\delta}_j \right) + \frac{\delta_j - \overline{\delta}_j}{\overline{\delta}_j \ln 2} \leq \mu_j, \quad j \in \mathcal{S}. \label{eq:P3-w-mu}
\end{align}

The optimization problem \eqref{eq:P3} for a fixed $\overline{\mathbf{V}}$ becomes a convex problem, given by 
\begin{subequations}
\label{eq:P3-w}
\begin{align}
\max_{\mathbf{R}_s, \mathbf{W}, \mathbf{r}, \alpha, \beta, \delta, \mu, \omega} \quad & \omega \\
\text{s.t.} \quad \quad \quad
& \eqref{eq:P3-w-a-c}, \eqref{eq:P3-w-a-p}, \eqref{eq:P3-w-mu}, \eqref{eq:P3-b-c}, \eqref{eq:P3-b-p}, \eqref{eq:P3-w-delta}, \eqref{eq:P3-g}, \\
& \eqref{eq:P2-r}, \eqref{eq:P2-r-p}, \eqref{eq:P2-r-c}, \eqref{eq:P1-r-c}, \eqref{eq:P1-p}.
\end{align}
\end{subequations}
Since \eqref{eq:P3-w} is convex problem, standard solvers such as CVXPY \cite{diamond2016cvxpy} can be utilized to obtain the solution efficiently.

\subsection{Optimization of STAR-RIS Coefficients}
In this step, the set of variables $\overline{\mathbf{R}}_s, \overline{\mathbf{W}}$ are fixed at each iteration step $\iota$ to optimize the remaining variables. 
The problem \eqref{eq:P3} is equivalent rewritten to the following problem
\begin{subequations}
\label{eq:P3-c}
\begin{align}
\max_{\mathbf{V}, \mathbf{r}, \alpha, \beta, \delta, \mu, \omega } \quad & \omega \\
\text{s.t.} \quad \quad
& \tilde{R}_{c, k}(\overline{\mathbf{R}}_s, \overline{\mathbf{W}}, {\mathbf{V}}) \geq \alpha_{c}, \forall k \in \mathcal{C}, \label{eq:P3-c-a-c} \\
& \tilde{R}_{p, k}(\overline{\mathbf{R}}_s, \overline{\mathbf{W}}, {\mathbf{V}}) \geq \alpha_{p, k}, \forall k \in \mathcal{C}, \label{eq:P3-c-a-p} \\
& D_j (\overline{\mathbf{R}}_s, \overline{\mathbf{W}}) \geq \delta_j, \forall j \in \mathcal{S}, \label{eq:P3-c-d} \\
& \eqref{eq:P3-w-mu}, \eqref{eq:P3-w-mu}, \eqref{eq:P3-b-c}, \eqref{eq:P3-b-p}, \eqref{eq:P3-g}, \eqref{eq:P2-r},  \\
& \eqref{eq:P2-r-p}, \eqref{eq:P2-r-c}, \eqref{eq:P1-r-c}, \eqref{eq:P1-c}, \eqref{eq:P1-c-1}.
\end{align}
\end{subequations}

The final problem formulation is convex, it can be efficiently solved using CVXPY \cite{diamond2016cvxpy}. The optimization procedure is summarized in Algorithm \ref{Alg:2}.

\subsection{Optimization of Conventional RIS Coefficients}
In this section, we further consider the case of optimizing the coefficients of a conventional RIS. Based on problem \eqref{eq:P3-c}, two additional conditions can be incorporated to enable the optimization for the conventional RIS case as follows.
\begin{subequations}
\label{eq:P3-c-ms}
\begin{align}
\max_{\mathbf{V}, \mathbf{r}, \alpha, \beta, \delta, \mu, \omega } \quad & \omega \\
\text{s.t.} \quad \quad
& {v}_{t, n} = 0, \quad \forall n \in \{1, 2, \dots, N_s/2 \}, \\
& {v}_{r, n} = 0, \; \, \forall n \in \{N_s/2 + 1, ..., N_s \} , \\
& \eqref{eq:P3-c-a-c}, \eqref{eq:P3-c-a-p}, \eqref{eq:P3-c-d} \\
& \eqref{eq:P3-w-mu}, \eqref{eq:P3-w-mu}, \eqref{eq:P3-b-c}, \eqref{eq:P3-b-p}, \eqref{eq:P3-g}, \eqref{eq:P2-r},  \\
& \eqref{eq:P2-r-p}, \eqref{eq:P2-r-c}, \eqref{eq:P1-r-c}, \eqref{eq:P1-c}, \eqref{eq:P1-c-1}.
\end{align}
\end{subequations}

\begin{algorithm}[htbp]
\caption{Iterative Algorithm for Maximizing the Minimum Total Secrecy Rate.}
\begin{algorithmic}
\label{Alg:2}
{\small
\STATE \textbf{Initialization}: \\
Set the iteration index $\iota  = 0$, a tolerance $\epsilon>0$, and a maximum number of iterations $\iota_{\sf{max}}$. \\
Generate an initial feasible point $\overline{\mathbf{R}}_s, \overline{\mathbf{W}}, \overline{\mathbf{V}}$.
\REPEAT
\STATE Solve the convex problem \eqref{eq:P3-w} to update $ \overline{\mathbf{R}}_s, \overline{\mathbf{W}}$;
\STATE Solve the convex problem \eqref{eq:P3-c} to update $ \overline{\mathbf{V}}, \omega^{(\iota + 1)}$;
\STATE Compute $\Delta {\omega}  = | {\omega^{\left(\iota + 1\right)} - \omega^{\left({\iota}\right)}} |$;
\STATE Update $\iota  \leftarrow \iota  + 1$;
\UNTIL the condition {$  {\Delta \omega } \leq \epsilon$} is satisfied or $\iota = \iota_{\sf{max}}$; \\
\STATE \textbf{Output}: \\
Return the optimized BS beamforming matrix and STAR-RIS coefficients, $\{ \overline{\mathbf{R}}_s, \overline{\mathbf{W}},\overline{\mathbf{V}}\}$.
}
\end{algorithmic}
\end{algorithm}

\section{Numerical Results}

This section presents simulation results along with an ablation study to evaluate the performance of the proposed algorithm in enhancing the security of the STAR-RIS-assisted RSMA-ISAC system. We employ a two-dimensional coordinate setup, where the BS and STAR-RIS are located at (0, 0) and (30m, 30m), respectively. Two sensing targets are located at the angle of  $30^{\circ}$ and  $-30^{\circ}$ from the BS while six CUs are randomly distributed at the maximum distance of 10m from STAR-RIS, three for reach region. All channels are modeled using Rician fading with a K-factor of 5dB. The large-scale fading is configured with a reference path loss of  $10^{-3}$ at 1m and a path loss exponent of 2.2 for all links, except for the BS-CU link, which use an exponent of 4.0. Other key parameters are set as following, if no otherwise specified. $P_B^{\max} = 30$ dBm, $N_B=8, N_{S}=32, \sigma_{c, k}^2 =\sigma_{s, k}^2 = -80$ dBm, $\forall k$, $\eta_k = -1$ dB, $\forall k$ , $\iota_{\max} = 20, \epsilon = 10^{-4}$.

Performance of proposed scheme is compared with several baselines. These include STAR-RIS ISAC with SDMA, RSMA ISAC with randomly configured STAR-RIS, conventional RIS and no RIS. The evaluation covers a wide range of scenarios, including different numbers of BS antennas, users, and STAR-RIS elements, as well as varying beampattern gain requirements to reflect sensing constraints. All simulation results are averaged over 200 independent random channel realizations using Monte Carlo simulations.

\begin{figure}[htbp]
    \centering
    \includegraphics[width=1\linewidth]{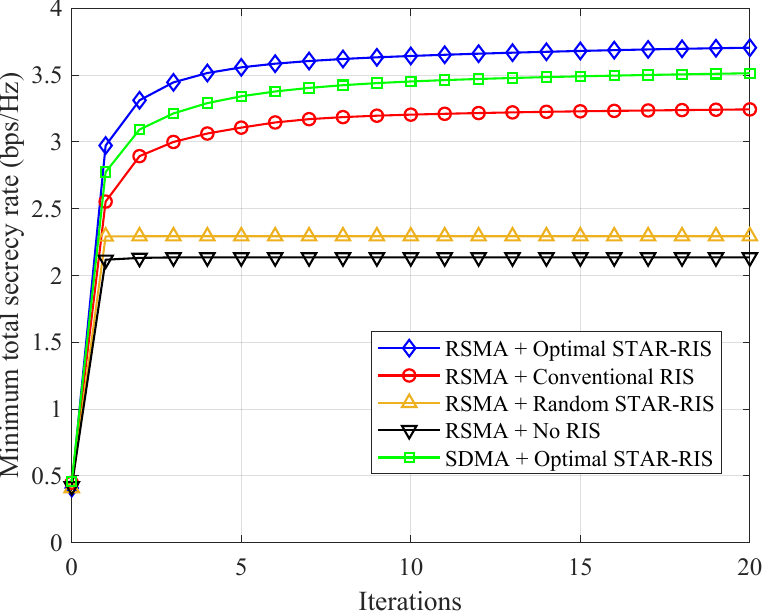}
    \caption{Convergence of the iterative algorithm for optimizing minimum total secrecy rate.}
    \label{fig: problem2-conv}
\end{figure}

We first examine the convergence behavior of the proposed algorithm. As shown in Fig.~\ref{fig: problem2-conv}, the gain increases rapidly and converges after approximately 3 iterations, demonstrating fast and stable convergence, making it highly suitable for practical implementation where computational efficiency is critical. Furthermore, the  RSMA with Optimal STAR-RIS proposed scheme is superior over others on the minimum secrecy rate. These results collectively demonstrate the effectiveness of the proposed algorithm with respect to both convergence speed and secrecy rate performance.

\begin{figure*}[ht]
    \centering
    \begin{subfigure}[t]{0.24\textwidth}
        \centering
        \includegraphics[width=\linewidth]{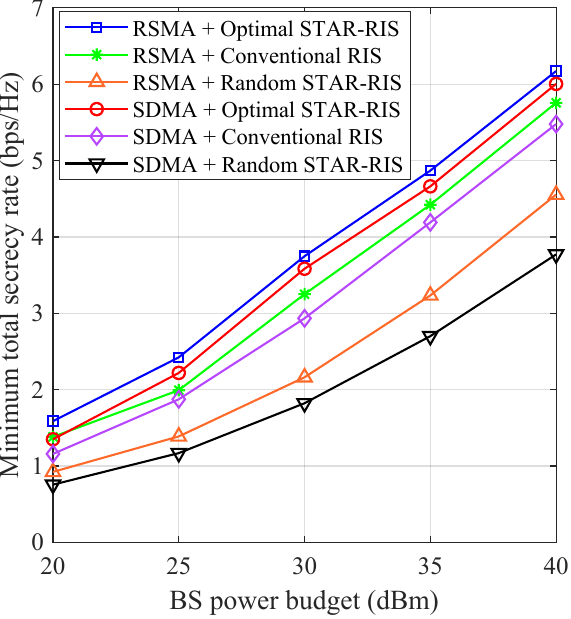}
        \caption{}
        \label{fig:scan-power}
    \end{subfigure}
    \hfill
    \begin{subfigure}[t]{0.248\textwidth}
        \centering
        \includegraphics[width=\linewidth]{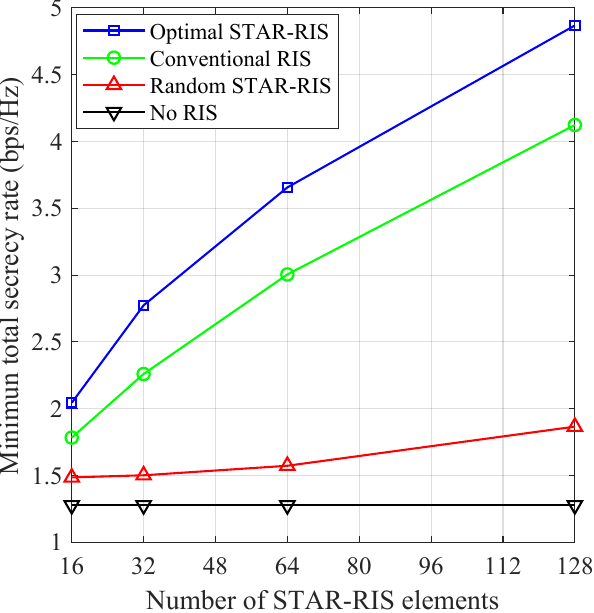}
        \caption{}
        \label{fig:scan-star}
    \end{subfigure}
    \hfill
    \begin{subfigure}[t]{0.24\textwidth}
        \centering
        \includegraphics[width=\linewidth]{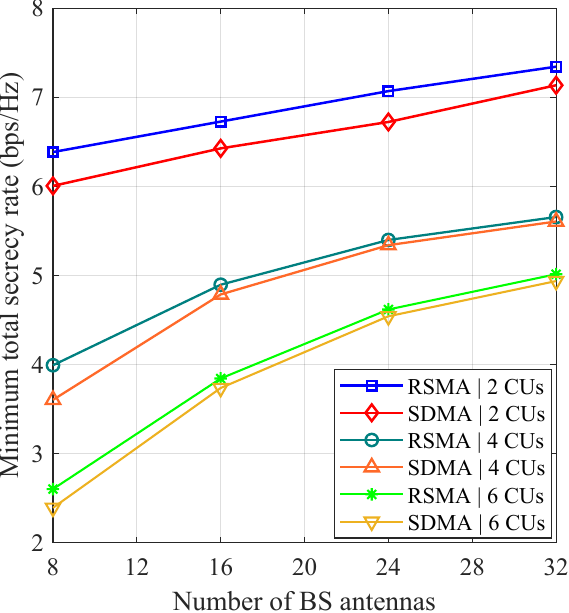}
        \caption{}
        \label{fig:scan-nbs}
    \end{subfigure}
    \hfill
    \begin{subfigure}[t]{0.245\textwidth}
        \centering
        \includegraphics[width=\linewidth]{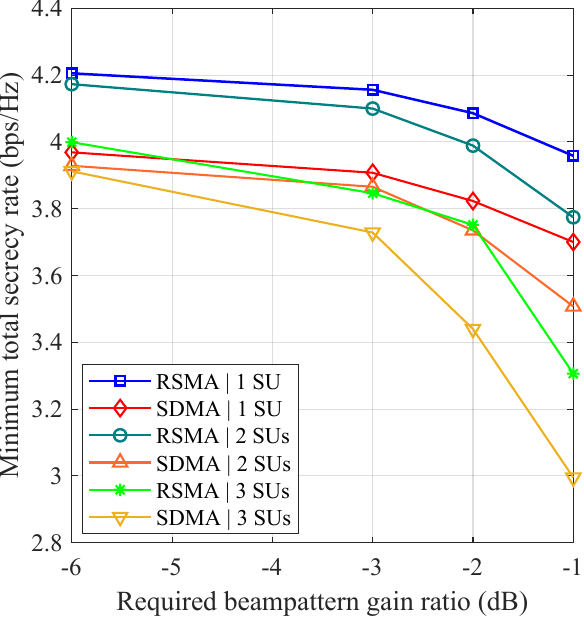}
        \caption{}
        \label{fig:scan-gain}
    \end{subfigure}
    \caption{Secrecy performance versus BS power budget, the number of STAR-RIS elements, the number of BS antennas and required beampattern gain under different system designs and scenarios.}
    \label{fig:scan-combined}
\end{figure*}

Fig.~\ref{fig:scan-power} shows the influence of the BS transmit power budget on the minimum total secrecy rate under various RIS configurations and multiple access schemes. As the power budget increases, all schemes exhibit improved secrecy performance. RSMA with optimized STAR-RIS consistently achieves the highest performance across all power levels, confirming its effectiveness in interference management and RIS utilization. Compared to SDMA, RSMA outperforms under every RIS setting. Moreover, optimized STAR-RIS provides clear advantages over conventional and random configurations, especially at higher power budgets, where the performance gap becomes more pronounced. These results emphasize the necessity of joint BS and RIS optimization to fully exploit the benefits of RSMA in secure communication.
 
Regarding the impact of RIS's configuration, Fig.~\ref{fig:scan-star} highlights that the secrecy rate is lowest without RIS and improves when STAR-RIS is applied. The random STAR-RIS shows limited gain as the number of elements increases. In contrast, the optimal STAR-RIS significantly boosts secrecy rate and scales efficiently with more elements. The conventional RIS provides a balance, it enhances performance better than random configuration and is simpler in hardware and more energy-efficient than the optimal one. These results underscore that while increasing the number of elements contributes to improved performance, intelligent configuration of the STAR-RIS is crucial to fully exploit its potential in secure communication systems.

Next, we investigate the the impact of multiple access technology selection and the number of CUs on secrecy rate. As illustrated in Fig.~\ref{fig:scan-nbs}, RSMA consistently outperforms SDMA on throughout several configurations of different numbers of BS antennas and CUs, verifying the effectiveness of interference mitigating mechanism of RSMA. As the number of antennas increases, the secrecy performance improves for all cases. In contrast, increasing the number of users leads to a reduction in secrecy rate due to higher interference. Notably, the performance gap becomes more pronounced with fewer antennas and fewer users, highlighting RSMA’s superior interference management capabilities. However, as the number of antennas increases, the performance gap between RSMA and SDMA gradually narrows, suggesting both schemes tend to converge in high-antenna regimes.

Finally, the trade-off capability between secure communication and sensing performance is studied in Fig.~\ref{fig:scan-gain}. As the required beampattern gain increases, which indicates enhanced sensing quality, the minimum total secrecy rate gradually decreases due to the resource shift toward sensing functions. In addition, increasing the number of sensing users from two to three further reduces the secrecy rate, revealing the resource competition between sensing and communication. Despite these challenges, RSMA consistently outperforms SDMA across all configurations, demonstrating its superior capability in balancing conflicting requirements of joint communication and sensing, even when facing with the rising number of SUs.

\section{Conclusions}
This paper investigated secure communication in ISAC systems assisted by STAR-RIS and RSMA, where the base station beamforming vectors and STAR-RIS coefficients were jointly optimized to enhance security in a fairness-aware manner while ensuring a minimum level of sensing performance. An iterative algorithm based on AO and MM was proposed to efficiently address the non-convex problem. Simulation results confirmed the rapid convergence and substantial performance gains enabled by the optimized STAR-RIS configuration. RSMA consistently outperformed SDMA in the presence of multiple sensing targets, which is more noticeable in scenarios with limited antennas or users. Furthermore, increasing the number of STAR-RIS elements significantly improved security performance, but only when their coefficients were optimally configured. The results demonstrated the superior capability of RSMA and STAR-RIS in collaboratively managing the trade-off between sensing and security, highlighting their potential for future secure ISAC networks.

\section*{Acknowledgment}
We would like thank Viettel High Technology Industries Corporation, Viettel Group for providing essential resources enabling the completion of this work.

\bibliographystyle{IEEEtran}
\bibliography{references} 

\end{document}